\pdfoutput=1

\documentclass[12pt,a4paper]{article}

\usepackage{ifthen} 
\newboolean{pdflatex}
\setboolean{pdflatex}{true} 

\newboolean{articletitles}
\setboolean{articletitles}{true} 

\newboolean{uprightparticles}
\setboolean{uprightparticles}{false} 

\newboolean{inbibliography}
\setboolean{inbibliography}{false} 

\usepackage{microtype}
\usepackage{lineno}  
\usepackage{xspace} 
\usepackage{caption} 

\usepackage{graphicx}  
\usepackage{color}
\usepackage{colortbl}
\graphicspath{{./figs/}} 

\usepackage{amsmath} 
\usepackage{amssymb}
\usepackage{amsfonts}
\usepackage{upgreek} 

\newcommand*\patchAmsMathEnvironmentForLineno[1]{%
\expandafter\let\csname old#1\expandafter\endcsname\csname #1\endcsname
\expandafter\let\csname oldend#1\expandafter\endcsname\csname
end#1\endcsname
 \renewenvironment{#1}%
   {\linenomath\csname old#1\endcsname}%
   {\csname oldend#1\endcsname\endlinenomath}%
}
\newcommand*\patchBothAmsMathEnvironmentsForLineno[1]{%
  \patchAmsMathEnvironmentForLineno{#1}%
  \patchAmsMathEnvironmentForLineno{#1*}%
}
\AtBeginDocument{%
\patchBothAmsMathEnvironmentsForLineno{equation}%
\patchBothAmsMathEnvironmentsForLineno{align}%
\patchBothAmsMathEnvironmentsForLineno{flalign}%
\patchBothAmsMathEnvironmentsForLineno{alignat}%
\patchBothAmsMathEnvironmentsForLineno{gather}%
\patchBothAmsMathEnvironmentsForLineno{multline}%
\patchBothAmsMathEnvironmentsForLineno{eqnarray}%
}

\usepackage{hyperref}    
\usepackage[all]{hypcap} 

\usepackage{xspace} 
\usepackage{upgreek}

\def\lhcb {\mbox{LHCb}\xspace}

\def\dzero  {\mbox{D0}\xspace}

\def\MagUp {\mbox{\em Mag\kern -0.05em Up}\xspace}

\ifthenelse{\boolean{uprightparticles}}%
{

 \def\Pmu         {\ensuremath{\upmu}\xspace}                 
 \def\Pnu         {\ensuremath{\upnu}\xspace}                 
                  
 \def\Ppi         {\ensuremath{\uppi}\xspace}

 \def\Pphi        {\ensuremath{\upphi}\xspace}

 \def\Ppsi        {\ensuremath{\uppsi}\xspace}

 \def\PDelta      {\ensuremath{\Delta}\xspace}                 
 \def\PXi      {\ensuremath{\Xi}\xspace}                 
 \def\PLambda      {\ensuremath{\Lambda}\xspace}                 
 \def\PSigma      {\ensuremath{\Sigma}\xspace}                 
 \def\POmega      {\ensuremath{\Omega}\xspace}                 
 \def\PUpsilon      {\ensuremath{\Upsilon}\xspace}                 
 
 \def\PB      {\ensuremath{\mathrm{B}}\xspace}                 
                  
 \def\PD      {\ensuremath{\mathrm{D}}\xspace}

 \def\PJ      {\ensuremath{\mathrm{J}}\xspace}                 
 \def\PK      {\ensuremath{\mathrm{K}}\xspace}

 \def\PX      {\ensuremath{\mathrm{X}}\xspace}

 \def\Pb      {\ensuremath{\mathrm{b}}\xspace}                 
 \def\Pc      {\ensuremath{\mathrm{c}}\xspace}                 
 \def\Pd      {\ensuremath{\mathrm{d}}\xspace}

 \def\Pi      {\ensuremath{\mathrm{i}}\xspace}

 \def\Pp      {\ensuremath{\mathrm{p}}\xspace}

 \def\Ps      {\ensuremath{\mathrm{s}}\xspace}

}
{

 \def\Pmu         {\ensuremath{\mu}\xspace}                 
 \def\Pnu         {\ensuremath{\nu}\xspace}                 
                  
 \def\Ppi         {\ensuremath{\pi}\xspace}

 \def\Pphi        {\ensuremath{\phi}\xspace}

 \def\Ppsi        {\ensuremath{\psi}\xspace}                 
                  
 \mathchardef\PDelta="7101
 \mathchardef\PXi="7104
 \mathchardef\PLambda="7103
 \mathchardef\PSigma="7106
 \mathchardef\POmega="710A
 \mathchardef\PUpsilon="7107
                  
 \def\PB      {\ensuremath{B}\xspace}                 
                  
 \def\PD      {\ensuremath{D}\xspace}

 \def\PJ      {\ensuremath{J}\xspace}                 
 \def\PK      {\ensuremath{K}\xspace}

 \def\PX      {\ensuremath{X}\xspace}

 \def\Pb      {\ensuremath{b}\xspace}                 
 \def\Pc      {\ensuremath{c}\xspace}                 
 \def\Pd      {\ensuremath{d}\xspace}

 \def\Pi      {\ensuremath{i}\xspace}

 \def\Pp      {\ensuremath{p}\xspace}

 \def\Ps      {\ensuremath{s}\xspace}

}

\makeatletter
\ifcase \@ptsize \relax
  \newcommand{\miniscule}{\@setfontsize\miniscule{4}{5}}
\or
  \newcommand{\miniscule}{\@setfontsize\miniscule{5}{6}}
\or
  \newcommand{\miniscule}{\@setfontsize\miniscule{5}{6}}
\fi
\makeatother

\DeclareRobustCommand{\optbar}[1]{\shortstack{{\miniscule (\rule[.5ex]{1.25em}{.18mm})}
  \\ [-.7ex] $#1$}}

\def\mup        {{\ensuremath{\Pmu^+}}\xspace}
\def\mun        {{\ensuremath{\Pmu^-}}\xspace} 

\def\neu        {{\ensuremath{\Pnu}}\xspace}

\def\neum       {{\ensuremath{\neu_\mu}}\xspace}

\def\dquark    {{\ensuremath{\Pd}}\xspace}

\def\squark    {{\ensuremath{\Ps}}\xspace}

\def\cquark    {{\ensuremath{\Pc}}\xspace}
\def\cquarkbar {{\ensuremath{\overline \cquark}}\xspace}

\def\bquark    {{\ensuremath{\Pb}}\xspace}

\def\pion   {{\ensuremath{\Ppi}}\xspace}

\def\pip    {{\ensuremath{\pion^+}}\xspace}
\def\pim    {{\ensuremath{\pion^-}}\xspace}

\def\pimp   {{\ensuremath{\pion^\mp}}\xspace}

\def\kaon    {{\ensuremath{\PK}}\xspace}
  \def\Kbar    {{\kern 0.2em\overline{\kern -0.2em \PK}{}}\xspace}

\def\KorKbar    {\kern 0.18em\optbar{\kern -0.18em K}{}\xspace}

\def\Kp      {{\ensuremath{\kaon^+}}\xspace}
\def\Km      {{\ensuremath{\kaon^-}}\xspace}
\def\Kpm     {{\ensuremath{\kaon^\pm}}\xspace}
\def\Kmp     {{\ensuremath{\kaon^\mp}}\xspace}
\def\KS      {{\ensuremath{\kaon^0_{\mathrm{ \scriptscriptstyle S}}}}\xspace}

  \def\Dbar    {{\kern 0.2em\overline{\kern -0.2em \PD}{}}\xspace}
\def\D       {{\ensuremath{\PD}}\xspace}

\def\DorDbar    {\kern 0.18em\optbar{\kern -0.18em D}{}\xspace}
\def\Dz      {{\ensuremath{\D^0}}\xspace}
\def\Dzb     {{\ensuremath{\Dbar{}^0}}\xspace}

\def\Dm      {{\ensuremath{\D^-}}\xspace}

\def\Dstarp  {{\ensuremath{\D^{*+}}}\xspace}
\def\Dstarm  {{\ensuremath{\D^{*-}}}\xspace}

\def\Dsp     {{\ensuremath{\D^+_\squark}}\xspace}
\def\Dsm     {{\ensuremath{\D^-_\squark}}\xspace}

\def\Dsmp    {{\ensuremath{\D^{\mp}_\squark}}\xspace}

\def\B       {{\ensuremath{\PB}}\xspace}
\def\Bbar    {{\ensuremath{\kern 0.18em\overline{\kern -0.18em \PB}{}}}\xspace}
\def\Bb      {{\ensuremath{\Bbar}}\xspace}
\def\BorBbar    {\kern 0.18em\optbar{\kern -0.18em B}{}\xspace}
\def\Bz      {{\ensuremath{\B^0}}\xspace}

\def\Bu      {{\ensuremath{\B^+}}\xspace}
\def\Bub     {{\ensuremath{\B^-}}\xspace}
\def\Bp      {{\ensuremath{\Bu}}\xspace}
\def\Bm      {{\ensuremath{\Bub}}\xspace}

\def\Bd      {{\ensuremath{\B^0}}\xspace}
\def\Bs      {{\ensuremath{\B^0_\squark}}\xspace}
\def\Bsb     {{\ensuremath{\Bbar{}^0_\squark}}\xspace}

\def\jpsi     {{\ensuremath{{\PJ\mskip -3mu/\mskip -2mu\Ppsi\mskip 2mu}}}\xspace}

  \def\Y#1S{\ensuremath{\PUpsilon{(#1S)}}\xspace}

\def\proton      {{\ensuremath{\Pp}}\xspace}
\def\antiproton  {{\ensuremath{\overline \proton}}\xspace}

\def\Lz          {{\ensuremath{\PLambda}}\xspace}
\def\Lbar        {{\ensuremath{\kern 0.1em\overline{\kern -0.1em\PLambda}}}\xspace}
\def\LorLbar    {\kern 0.18em\optbar{\kern -0.18em \PLambda}{}\xspace}

\def\Lb      {{\ensuremath{\Lz^0_\bquark}}\xspace}

\def\Lcbar   {{\ensuremath{\Lbar{}^-_\cquark}}\xspace}

\def\to                 {\ensuremath{\rightarrow}\xspace}

\def\CP                {{\ensuremath{C\!P}}\xspace}
\def\CPT               {{\ensuremath{C\!PT}}\xspace}

\newcommand{\dm}{{\ensuremath{\Delta m}}\xspace}
\newcommand{\dms}{{\ensuremath{\Delta m_{\squark}}}\xspace}

\newcommand{\DGd}{{\ensuremath{\Delta\Gamma_{\dquark}}}\xspace}

\newcommand{\Gd}{{\ensuremath{\Gamma_{\dquark}}}\xspace}

\def\AT#1     {\ensuremath{A_{\mathrm{T}}^{#1}}\xspace}           

\def\C#1      {\ensuremath{\mathcal{C}_{#1}}\xspace}                       
\def\Cp#1     {\ensuremath{\mathcal{C}_{#1}^{'}}\xspace}                    
\def\Ceff#1   {\ensuremath{\mathcal{C}_{#1}^{\mathrm{(eff)}}}\xspace}        
\def\Cpeff#1  {\ensuremath{\mathcal{C}_{#1}^{'\mathrm{(eff)}}}\xspace}       
\def\Ope#1    {\ensuremath{\mathcal{O}_{#1}}\xspace}                       
\def\Opep#1   {\ensuremath{\mathcal{O}_{#1}^{'}}\xspace}                    

\newcommand{\tev}{\ifthenelse{\boolean{inbibliography}}{\ensuremath{~T\kern -0.05em eV}\xspace}{\ensuremath{\mathrm{\,Te\kern -0.1em V}}}\xspace}
\newcommand{\gev}{\ensuremath{\mathrm{\,Ge\kern -0.1em V}}\xspace}
\newcommand{\mev}{\ensuremath{\mathrm{\,Me\kern -0.1em V}}\xspace}
\newcommand{\kev}{\ensuremath{\mathrm{\,ke\kern -0.1em V}}\xspace}
\newcommand{\ev}{\ensuremath{\mathrm{\,e\kern -0.1em V}}\xspace}
\newcommand{\gevc}{\ensuremath{{\mathrm{\,Ge\kern -0.1em V\!/}c}}\xspace}
\newcommand{\mevc}{\ensuremath{{\mathrm{\,Me\kern -0.1em V\!/}c}}\xspace}
\newcommand{\gevcc}{\ensuremath{{\mathrm{\,Ge\kern -0.1em V\!/}c^2}}\xspace}
\newcommand{\gevgevcccc}{\ensuremath{{\mathrm{\,Ge\kern -0.1em V^2\!/}c^4}}\xspace}
\newcommand{\mevcc}{\ensuremath{{\mathrm{\,Me\kern -0.1em V\!/}c^2}}\xspace}

\def\invfb   {\ensuremath{\mbox{\,fb}^{-1}}\xspace}

\def\gsim{{~\raise.15em\hbox{$>$}\kern-.85em
          \lower.35em\hbox{$\sim$}~}\xspace}
\def\lsim{{~\raise.15em\hbox{$<$}\kern-.85em
          \lower.35em\hbox{$\sim$}~}\xspace}

\def\pt         {\mbox{$p_{\mathrm{ T}}$}\xspace}

\def\tell1  {TELL1\xspace}
\def\ukl1   {UKL1\xspace}

\usepackage{cite} 
\usepackage{mciteplus}

\usepackage{longtable} 
\def\myTitle {Measurement of the \CP asymmetry in \Bs--\Bsb mixing}

\def\aslsVal    {0.39}
\def\aslsStat   {0.26}
\def\aslsSyst   {0.20}
\def\arawval    {0.11}
\def\arawstat   {0.09}

\newcommand{\fbkgresult}{\ensuremath{\fbkg = (18.4 \pm 6.0)\%}\xspace}
\newcommand{\fbkgabkgresult}{\ensuremath{\fbkg \Abkg = (-0.023 \pm 0.031)\%}\xspace}

\def\asld   {\ensuremath{a_{\rm sl}^{d}}\xspace}
\def\asls   {\ensuremath{a_{\rm sl}^{s}}\xspace}
\def\asl    {\ensuremath{a_{\rm sl}}\xspace}
\def\fbar   {\ensuremath{\bar{f}}\xspace}

\usepackage{cancel}

\newcommand{\Araw}{\ensuremath{A_{\rm raw}}\xspace}
\newcommand{\Abkg}{\ensuremath{A_{\rm bkg}}\xspace}
\newcommand{\fbkg}{\ensuremath{f_{\rm bkg}}\xspace}
\newcommand{\Abkgi}{\ensuremath{A_{\rm bkg}^i}\xspace}
\newcommand{\fbkgi}{\ensuremath{f_{\rm bkg}^i}\xspace}

\newcommand{\Adet}{\ensuremath{A_{\rm det}}\xspace}

\def\amupitrack {{\ensuremath{A_{\text{track}}({\pim\mup})}}\xspace}
\def\akktrack {{\ensuremath{A_{\text{track}}({\Kp\Km})}}\xspace}
\def\atrack {{\ensuremath{A_{\text{track}}}}\xspace}
\def\apid {{\ensuremath{A_{\text{PID}}}}\xspace}
\def\amu {{\ensuremath{A_{\text{trig}}({\text{hardware}})}}\xspace}
\def\ahlt {{\ensuremath{A_{\text{trig}}({\text{software}})}}\xspace}
\def\atrig{{\ensuremath{A_{\text{trig}}}}\xspace}

\def\phipi  {{\ensuremath{\phi\pi}}\xspace}
\def\kstark {{\ensuremath{K^*K}}\xspace}
\def\kkpinr {{\ensuremath{\text{NR}}}\xspace}
\def\nr     {\kkpinr}

\def\Mcorr{{\ensuremath{m_{\text{corr}}}}\xspace}
\def\IP{{\ensuremath{\text{IP}}}\xspace}

\def\mupm        {{\ensuremath{\Pmu^{\pm}}}\xspace}

\def\numuoptbar {\shortstack{{\miniscule \scalebox{0.8}{(\rule[.5ex]{1.25em}{.18mm})}} \\ [-.7ex] \neu}_{\kern -0.4ex \mu}\xspace}

\DeclareGraphicsExtensions{.pdf,.PDF,png,.PNG}

\newboolean{wordcount}
\setboolean{wordcount}{false} 

\def\figWidth{0.65}

\begin{document}

\renewcommand{\thefootnote}{\fnsymbol{footnote}}
\setcounter{footnote}{1}


\begin{titlepage}
\pagenumbering{roman}

\vspace*{-1.5cm}
\centerline{\large EUROPEAN ORGANIZATION FOR NUCLEAR RESEARCH (CERN)}
\vspace*{1.5cm}
\noindent
\begin{tabular*}{\linewidth}{lc@{\extracolsep{\fill}}r@{\extracolsep{0pt}}}
\ifthenelse{\boolean{pdflatex}}
{\vspace*{-2.7cm}\mbox{\!\!\!\includegraphics[width=.14\textwidth]{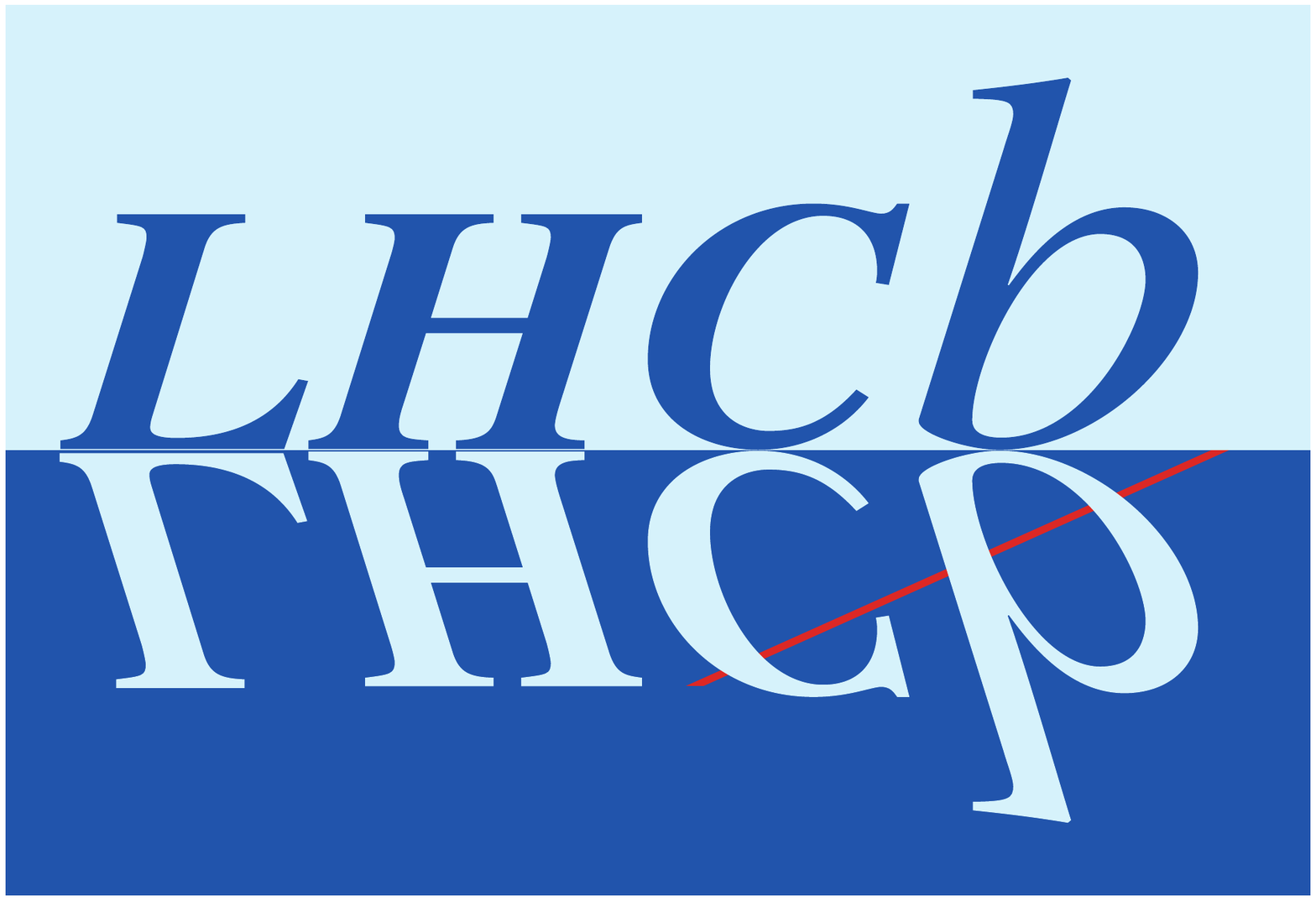}} & &}%
{\vspace*{-1.2cm}\mbox{\!\!\!\includegraphics[width=.12\textwidth]{lhcb-logo.eps}} & &}%
\\
 & & CERN-EP-2016-133 \\  
 & & LHCb-PAPER-2016-013 \\  
 & & 8 August 2016 \\
\end{tabular*}

\vspace*{3.0cm}

{\normalfont\bfseries\boldmath\huge
\begin{center}
  \myTitle
\end{center}
}

\vspace*{1.5cm}

\begin{center}
The LHCb collaboration\footnote{Authors are listed at the end of this paper.}
\end{center}

\vspace{\fill}

\begin{abstract}
  \noindent
  The $\CP$ asymmetry in the mixing of $\Bs$ and $\Bsb$ mesons is measured in proton--proton
collision data corresponding to an integrated luminosity of $3.0\invfb$,
recorded by the LHCb experiment at centre-of-mass energies of 7 and 8\tev.
Semileptonic $\Bs$ and $\Bsb$ decays are studied in the inclusive mode 
$\Dsmp \mupm \numuoptbar \PX$
with the $\Dsmp$ mesons reconstructed in the $\Kp \Km \pimp$ final state. 
Correcting the observed charge asymmetry for detection and
background effects, the \CP asymmetry is found to be $\asls = (\aslsVal \pm
\aslsStat \pm \aslsSyst)\%$, where the first uncertainty is statistical and the
second systematic. This is the most precise measurement of \asls to date. It is consistent with 
the prediction from the Standard Model and will constrain new models of particle physics.

\end{abstract}

\vspace*{2.0cm}

\begin{center}
  Published in Phys.~Rev.~Lett. {\bf 117} (2016), 061803
\end{center}

\vspace{\fill}

{\footnotesize 
\centerline{\copyright~CERN on behalf of the \lhcb collaboration, licence \href{http://creativecommons.org/licenses/by/4.0/}{CC-BY-4.0}.}}
\vspace*{2mm}

\end{titlepage}


\newpage
\setcounter{page}{2}
\mbox{~}
%
%
%
%

\cleardoublepage


\renewcommand{\thefootnote}{\arabic{footnote}}
\setcounter{footnote}{0}


\pagestyle{plain} 
\setcounter{page}{1}
\pagenumbering{arabic}


\noindent
When neutral \B mesons evolve in time they can change into their own
antiparticles. This quantum-mechanical phenomenon is known as mixing and occurs
in both neutral \B meson systems, \Bd and \Bs, where \B is used to refer to
either system. In this mixing process, the \CP (charge-parity) symmetry is
broken if the probability for a \B meson to change into a \Bbar meson is
different from the probability for the reverse process.  This effect can be
measured by studying decays into flavour-specific final states, $\B \to f$, such
that $\Bb \to f$ transitions can only occur through the mixing process $\Bb \to
\B \to f$.  Such processes include semileptonic \B decays, as the charge of the
lepton identifies the flavour of the \B meson at the time of its decay. The
magnitude of the \CP-violating asymmetry in \B mixing can be characterized by
the semileptonic asymmetry \asl. This is defined in terms of the partial decay
rates, $\Gamma$, to semileptonic final states as
\begin{equation}
  \asl \equiv \frac{\Gamma(\Bb \to f) - \Gamma(\B \to \fbar)}
       {\Gamma(\Bb \to f) + \Gamma(\B \to \fbar)} \approx 
       \frac{\Delta \Gamma}{\dm} \tan\phi_{12} \ , 
  \label{eq:asl}
\end{equation}
where \dm ($\Delta \Gamma$) is the difference in mass (decay width) between the
mass eigenstates of the $\B$ system and $\phi_{12}$ is a \CP-violating phase
\cite{Lenz:2006hd}. In the Standard Model (SM), the asymmetry is predicted to
be as small as $\asld = (-4.7\pm0.6)\times 10^{-4}$ in the \Bz system and $\asls
= (2.22\pm0.27)\times 10^{-5}$ in the \Bs system~\cite{Lenz:2006hd,
  Artuso:2015swg}. However, these values may be enhanced by
non-SM contributions to the mixing process~\cite{Lenz:2012NP}.

Measurements of \asl have led to an inconclusive picture. In 2010, the \dzero
collaboration reported an anomalous charge asymmetry in the inclusive production
rates of like-sign dimuons~\cite{Abazov:2010V1}, which is sensitive to a
combination of \asld and \asls. Their most recent study shows a discrepancy with
SM predictions of about three standard deviations~\cite{Abazov:2013uma}. The
current experimental world averages, excluding the anomalous \dzero result, are
$\asld = (0.01 \pm 0.20)\%$ and $\asls = (-0.48 \pm 0.48)\%$~\cite{HFAG},
compatible with both the SM predictions and the \dzero measurement. The
measurement of \asls presented in this letter is based on data recorded by \lhcb
in 2011 and 2012, corresponding to an integrated luminosity of 3.0\invfb. It
supersedes the previous \lhcb measurement~\cite{LHCb-PAPER-2013-033}, which used
the 1.0\invfb data sample taken in 2011. Semileptonic decays
$\Bs\to\Dsm\mup\neum\PX$, where \PX represents any number of particles, are
reconstructed inclusively in $\Dsm\mup$. Charge-conjugate modes are implied
throughout, except in the definitions of charge asymmetry. The \Dsm meson is
reconstructed in the $\Kp\Km\pim$ final state. This analysis extends the
previous \lhcb measurement, which considered only $\Dsm\to\phi\pim$ decays, by
including all possible \Dsm decays to the $\Kp\Km\pim$ final state.

Starting from a sample with equal numbers of \Bs and \Bsb mesons, \asls can be measured without determining (tagging) the initial flavour.
The raw asymmetry of observed \Dsm\mup and \Dsp\mun candidates, integrated over \Bs decay time, is
\begin{align}
 \Araw = \frac{N(\Dsm\mup) - N(\Dsp\mun)}{N(\Dsm\mup) + N(\Dsp\mun)} \ . 
\end{align}
The high oscillation frequency \dms reduces the effect of the small asymmetry in the 
production rates between \Bs and \Bsb mesons in $pp$ collisions by a
factor $10^{-3}$~\cite{LHCb-PAPER-2013-033, LHCb-PAPER-2014-042}.
Neglecting corrections, the untagged, time-integrated asymmetry is 
$\Araw=\asls/2$, where the factor two reduction compared to the tagged
asymmetry in Eq.~\ref{eq:asl} comes from the summation over 
mixed and unmixed decays.
The tagged asymmetry would actually suffer from a larger
reduction because of the tagging efficiency~\cite{LHCb-PAPER-2011-027, 
LHCb-PAPER-2015-056}. 
The unmixed decays have zero asymmetry due to \CPT symmetry.
The raw asymmetry is still affected
by possible differences in detection efficiency for the two charge-conjugate
final states and by backgrounds from other \bquark-hadron decays to $\Dsm \mup
\PX$. Hence, \asls is calculated as
\begin{align}
  \asls = \frac{2}{1-\fbkg} ( \Araw - \Adet - \fbkg \Abkg ) \ ,
  \label{eq:aslsExp}
\end{align}
where \Adet is the detection asymmetry, which is assessed from data using
calibration samples, \fbkg is the fraction of \bquark-hadron background and
\Abkg the background asymmetry.

The \lhcb detector is a single-arm forward spectrometer designed for the study
of particles containing \bquark or \cquark
quarks~\cite{Alves:2008zz,LHCb-DP-2014-002}. A high-precision tracking system
with a dipole magnet measures the momentum ($p$) and impact parameter (\IP) of
charged particles. The \IP is defined as the distance of closest approach
between the track and any primary proton--proton interaction and is used
to distinguish between \Dsm mesons from \B decays and \Dsm mesons promptly
produced in the primary interaction. The regular reversal of the magnet polarity
allows a quantitative assessment of detector-induced charge
asymmetries. Different types of charged particles are distinguished using
particle identification (PID) information from two ring-imaging Cherenkov
detectors, an electromagnetic calorimeter, a hadronic calorimeter and a muon
system.  Online event selection is performed by a two-stage trigger. For this analysis, the first
(hardware) stage selects muons in the muon system; the second (software) stage
applies a full event reconstruction. Here the events are first selected by the
presence of the muon or one of the hadrons from the \Dsm decay, after which a
combination of the decay products is required to be consistent
with the topological signature of a \bquark-hadron decay.  Simulated events are
produced using the software described in
Refs.~\cite{Sjostrand:2006za,*Sjostrand:2007gs,LHCb-PROC-2010-056, Lange:2001uf,
  Allison:2006ve, *Agostinelli:2002hh, LHCb-PROC-2011-006}.

\begin{figure}
  \centering
  \includegraphics[width=\figWidth\textwidth]{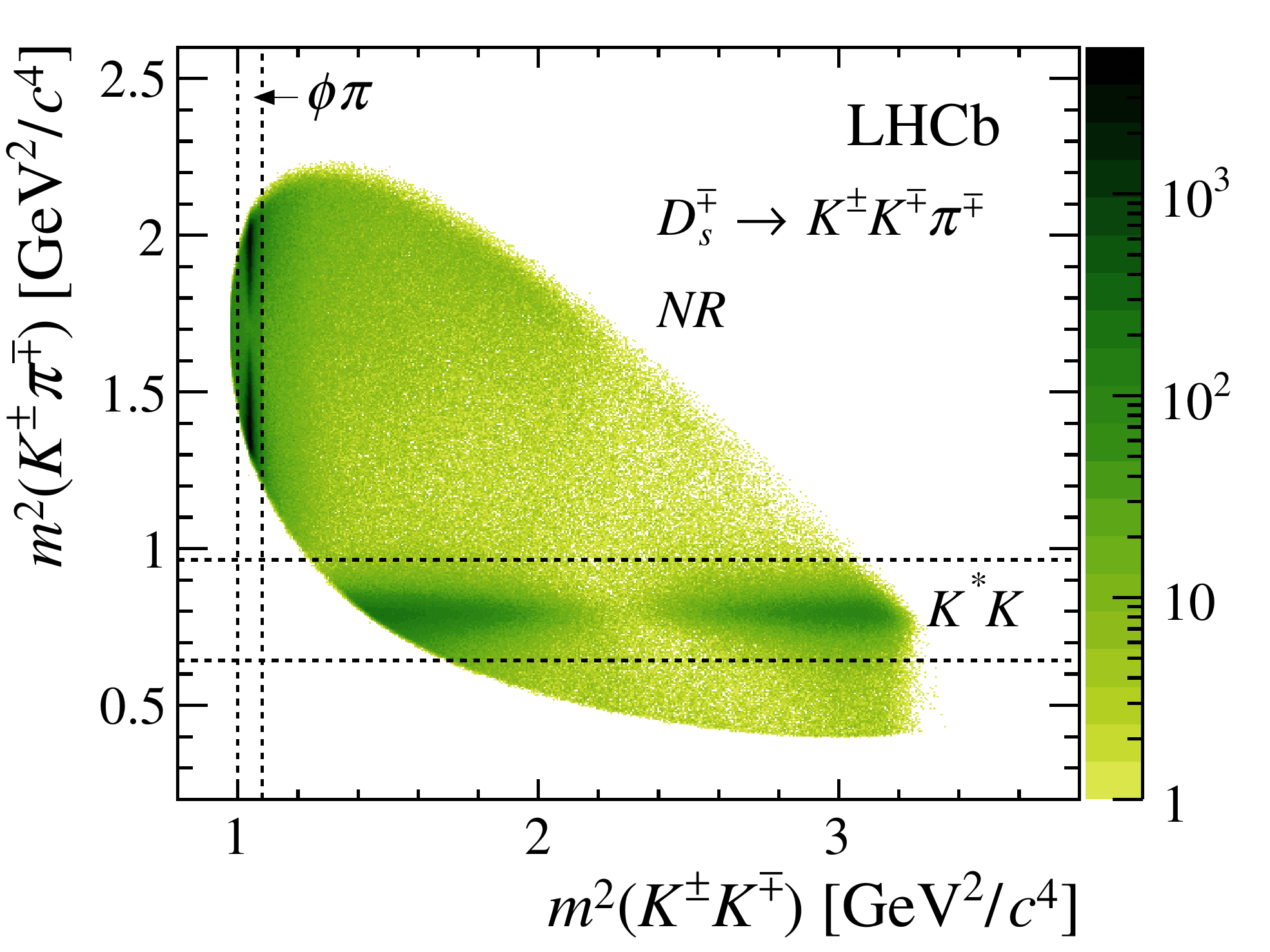}
  \caption{Dalitz plot of the $\Dsmp\to\Kpm\Kmp\pimp$ decay for selected
    $\Dsmp\mu^{\pm}$ candidates, with the three selection regions indicated. To
    suppress combinatorial background, a narrow invariant mass window, between
    1950 and 1990\mevcc, is required for the \Dsmp candidates in this plot.}
  \label{fig:DalitzRegions}
\end{figure}

Different intermediate states, clearly visible in the Dalitz plot shown 
in Fig.~\ref{fig:DalitzRegions}, contribute to the three-body $\Dsm\to\Kp\Km\pim$ decays. 
Three disjoint regions are defined, which have different levels of background.
The \phipi region is the cleanest and is selected by requiring the
reconstructed \Kp\Km mass to be within $\pm20\mevcc$ of the known \Pphi
mass. The \kstark region is selected by requiring the reconstructed \Kp\pim mass
to be within $\pm90\mevcc$ of the known $K^*(892)^0$ mass. The remaining \Dsm
candidates are included in the non-resonant (\kkpinr) region, which also covers
other intermediate states~\cite{PDG2014}.

The \Dsm candidates are reconstructed from three charged tracks, and then a muon
track with opposite charge is added. All four tracks are required to have a good
quality track fit and significant \IP. The contribution from prompt \Dsm
background is suppressed to a negligible level by imposing a lower bound on the
\IP of the \Dsm candidates. To ensure a good overlap with the calibration
samples, minimum momenta of 2, 5 and 6\gevc and minimum transverse momenta, \pt,
of 300, 400 and 1200\mevc are required for the pions, kaons and muons,
respectively. To suppress background, kaon and pion candidates are required to
be positively identified by the PID system. Candidates are selected by
requiring a good quality of the \Dsm and \Bs decay vertices. 
A source of background arises from \Dsm candidates where one of the three decay particles is
misidentified. The main contributions are from $\Lcbar \to \Kp \antiproton \pim$, $\Dm \to \Kp \pim \pim$, $\jpsi X$, and
misidentified or partially reconstructed multibody \D decays, all originating
from semileptonic \bquark-hadron decays. They are suppressed to a negligible level by specific
vetoes, which apply tight PID requirements in a small window of invariant mass of 
the corresponding particle combination. These vetoes are optimized separately for each Dalitz plot region.
To check
that this does not introduce additional asymmetries, these selections are
applied to control samples of promptly produced \Dsm mesons. The asymmetries are found to be consistent between the Dalitz regions.

The \Dsm\mup signal yields are obtained from fits to the $\Kp\Km\pim$ invariant
mass distributions. 
These yields contain contributions from backgrounds that also peak at the \Dsm mass, originating from other \bquark-hadron
decays into \Dsm mesons and muons.
Simulation studies indicate that these
peaking backgrounds are mainly composed of \bquark-hadron decays to $\Dsm X_c
X$, where the \Dsm meson originates from a $\bquark\to\cquark\cquarkbar\squark$
transition, and $X_c$ is a charmed hadron decaying semileptonically.

An example of such a background is $\Bm\to \Dsm \Dzb \PX$.
Other,
smaller contributors are $\Bp\to\Dsm\Kp\mup\neum\PX$ and $\Bz\to\Dsm\KS
\mup\neum\PX$ decays. All of these peaking backgrounds have more missing
particles than the $\Bs\to\Dsm\mup\neum\PX$ signal decay. Their contribution is
reduced by requiring the corrected \Bs mass, defined as $\Mcorr \equiv \sqrt{m^2 + \pt^2} + \pt$, 
to be larger than $4200\mevcc$, where $m$ is the
$\Dsm\mup$ invariant mass and \pt the $\Dsm\mup$ momentum transverse to the line connecting the primary and \Bs decay vertices.

The estimates of \fbkg and \Abkg are based on known branching
fractions~\cite{PDG2014}, selection efficiencies and background asymmetries,
using a similar approach as in the previous
measurement~\cite{LHCb-PAPER-2013-033}. The reconstruction and selection
efficiencies of the backgrounds relative to the signal efficiency are determined
from simulation. The total background asymmetry is given by the sum of all
contributions as $\fbkg\Abkg \equiv \sum_i \fbkgi\Abkgi$. The background
asymmetries mainly originate from the production asymmetries of \bquark hadrons.
The production asymmetry between \Bp and \Bm mesons is $\Abkg(\Bu) = (-0.6 \pm
0.6)\%$, obtained from the observed asymmetry in $\Bp\to\jpsi\Kp$
decays~\cite{LHCb-PAPER-2014-032}, after correcting for the kaon detection
asymmetry and the direct \CP asymmetry~\cite{PDG2014}. For the \Bd background,
there are contributions from the production asymmetry and from
\asld~\cite{LHCb-PAPER-2014-053}. Both asymmetries are diluted when integrating
over the \Bd decay time, resulting in $\Abkg(\Bd) = (-0.18 \pm 0.13)\%$. The
production asymmetry in the \Lb backgrounds is estimated based on the combined
\CP and production asymmetry measured in $\Lb \to \jpsi p^+\Km$
decays~\cite{LHCb-PAPER-2015-032}. The direct \CP asymmetry in this decay mode
is estimated to be $(-0.6\pm0.3)\%$, using the measurements in
Ref.~\cite{LHCb-PAPER-2014-020} and the method proposed in
Ref.~\cite{DeBruyn:2014oga}. Subtracting this from the combined
asymmetry~\cite{LHCb-PAPER-2015-032} results in $\Abkg(\Lb) = (+0.5 \pm
0.8)\%$. The overall peaking background fraction is \fbkgresult and the
correction for the background asymmetry is \fbkgabkgresult.

\begin{figure}
  \centering
  \includegraphics[width=\figWidth\textwidth]{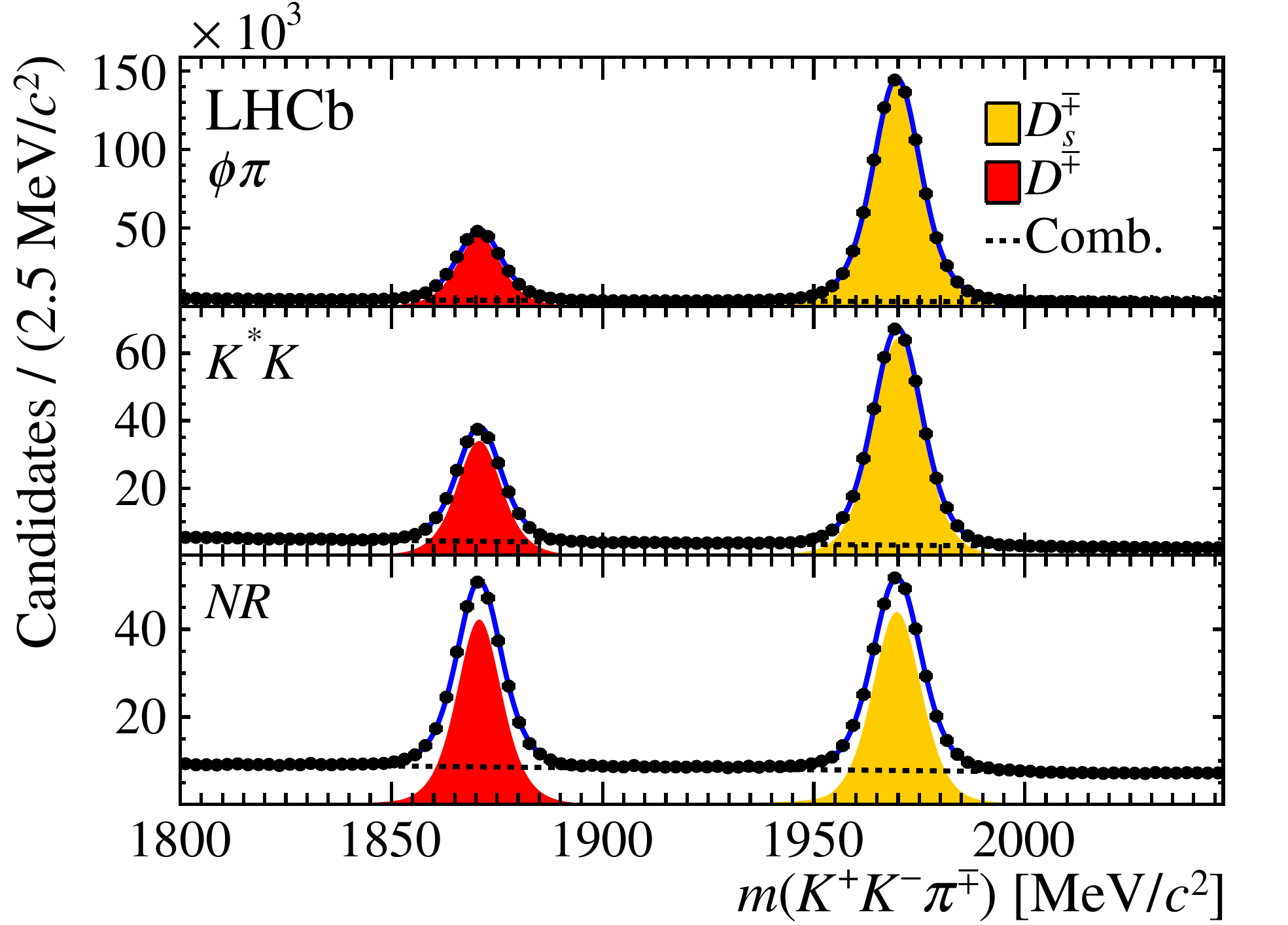}
  \caption{Distributions of $\Kp\Km\pimp$ mass in the three Dalitz plot regions,
    summed over both magnet polarities and data-taking periods. Overlaid is the
    result of the fit, with signal and combinatorial background components as
    indicated in the legend.}
  \label{fig:massfit}
\end{figure}

The $\Kp\Km\pimp$ mass distributions are shown in Fig.~\ref{fig:massfit}, with
the fit results superimposed. The \Dsmp\mupm yields are found to be
$899\times10^3$ in the \phipi region, $413\times10^3$ in the \kstark region, and
$280\times10^3$ in the \nr region. Extended maximum likelihood fits are
made separately for the three Dalitz regions, for the two magnet polarities, and
the two data-taking periods (2011 and 2012).  To accurately determine the
background shape from random combinations of $\Kp\Km\pim$ candidates, a wide
mass window between 1800 and 2047\mevcc is used, which includes the
Cabibbo-suppressed $\Dm \to \Kp \Km \pim$ decay. Both peaks are modelled with a
double-sided Hypatia function~\cite{Santos:2014}. The tail parameters of this
function are determined for each Dalitz region by a fit to the combined data
sets for all magnet polarities and data-taking periods, and subsequently fixed
in the twelve individual mass fits. A systematic uncertainty is assigned to account for
fixing these parameters. The combinatorial background is modelled with a
second-order polynomial. 
A simultaneous fit to the $m(\Kp\Km\pim)$ and
$m(\Kp\Km\pip)$ distributions is performed. All signal parameters except 
the mean masses and signal yields are shared between the \Dsm and \Dsp candidates.
All background parameters vary independently
in the fit to allow for any asymmetry in the combinatorial background. Possible
biases from the fit model are studied by generating invariant mass distributions
with the signal component described by a double Gaussian function with power-law
tails on both sides, and subsequently applying the fit with the default Hypatia
shape. The change in the value of \Araw is assigned as a systematic uncertainty.

Asymmetries are averaged as follows. For each magnet polarity and data-taking period, the weighted
average of the asymmetries of the three Dalitz regions is taken.
Then the arithmetic average for the
two magnet polarities is taken to minimize possible residual detection
asymmetries~\cite{LHCb-PAPER-2013-033}. Finally, a weighted average is made over the two data-taking
periods. The resulting raw asymmetry is $\Araw = (\arawval \pm \arawstat)\%$.


The asymmetry \Adet, arising from the difference in detection efficiencies
between the \Dsm\mup and \Dsp\mun candidates, is determined using calibration
samples. The asymmetry is split up as
\begin{equation}
\label{eq:Adet}
\Adet = \atrack + \apid + \atrig \ ,
\end{equation}
where the individual contributions are described below.  For each calibration
sample, event weights are applied to match the three-momentum distributions of
the calibration particles to those of the signal decays. The weights are
determined in bins of the distributions of momenta and angles.  Alternative
binning schemes are used to assess the systematic uncertainties due to the
weighting procedure.

The track reconstruction asymmetry, \atrack, is split into a contribution,
\akktrack, associated with the reconstruction of the $\Kp\Km$ pair and a contribution,
\amupitrack, associated with the $\pim\mup$ pair. The track reconstruction efficiency for
single kaons suffers from a sizeable difference between \Kp and \Km
cross-sections with the detector material, which depends on the kaon
momentum. This asymmetry largely cancels in \akktrack, due to the similar
kinematic distributions of the positive and negative kaons. The kaon asymmetry
is calculated using prompt $\Dm \to \Kp \pim \pim$ and $\Dm \to \KS \pim$
decays, similarly to Refs.~\cite{LHCb-PAPER-2014-013, LHCb-PAPER-2014-053}.
For pions and muons, the charge asymmetry due to interactions in the detector
material is assumed to be negligible, and a systematic uncertainty
is assigned for this assumption~\cite{LHCb-PAPER-2014-053}. Effects from the
track reconstruction algorithms and detector acceptance, combined with a
difference in kinematic distributions between pions and muons, can result in a
charge asymmetry. It is assessed here with two methods.  The first method
measures the track reconstruction efficiency using samples of partially
reconstructed $\jpsi \to \mup \mun$ decays as described in
Ref.~\cite{LHCb-DP-2013-002}. The second method uses fully and partially
reconstructed $\Dstarm \to \Dzb (\Kp \pim \pip \pim) \pim$ decays as described
in Ref.~\cite{LHCb-PAPER-2012-009}. The final value of \amupitrack is obtained
as the weighted average from the two methods. The systematic uncertainty on this
number includes a small effect from differences in the detector acceptance for
positive and negative particles.

The asymmetry induced by the PID requirements, \apid, is determined using large
samples of $\Dstarp \to \Dz (\Km \pip) \pip$ and $\jpsi\to\mup\mun$ decays. The
\Dstarp charge identifies the kaon and the pion of the \Dz decay without the
use of PID requirements, which is then used to determine the PID efficiencies
and corresponding charge asymmetries. 

The asymmetry induced by the trigger, \atrig, is split into contributions from
the muon hardware trigger and from the software trigger. The first, \amu, is
assessed using samples of $\jpsi \to \mup \mun$ decays in data. The second,
\ahlt, is mainly caused by the trigger requirements on the muon or one of the
hadrons from the \Dsm decay. The asymmetry from the muon software trigger is
determined in a similar fashion to that from the hardware trigger. The asymmetry
due to the trigger requirement on the hadrons is determined using samples of
prompt $\Dsm \to \Kp \Km \pim$ decays that have been triggered by other
particles in the event. The combined asymmetry takes into account the overlap
between the two triggers.

The measured values of all detection asymmetries with their
statistical and systematic uncertainties are shown in
Table~\ref{tab:systematics}. The overall corrections are small and compatible
with zero. In contrast, corrections for separate magnet polarities are more
significant (at most 1.1\% in 2011 and 0.3\% in 2012), as expected for
most of the detector-induced charge asymmetries. The corrections for the detection asymmetries are
almost fully correlated between the Dalitz regions.

\begin{table}
  \centering
  \caption{Overview of contributions in the determination of \asls, averaged
    over Dalitz plot regions, magnet polarities and data taking periods, with their
    statistical and systematic uncertainties. All numbers are in percent. The
    central value of \asls is calculated according to Eq.~\ref{eq:aslsExp}. The
    uncertainties are added in quadrature and multiplied by $2/(1-\fbkg)$,
    which is the same for all twelve subsamples, to obtain the uncertainties on
    \asls.}
  \label{tab:systematics} 
  \begin{tabular}{r|c c c l}
Source &  Value & Stat. uncert. & Syst. uncert. \\
\hline
\Araw                 & $\phantom{-}$\arawval & \arawstat & $0.02$ \\
$-$\akktrack          & $\phantom{-}0.01$ & $0.00$ & $0.03$ \\
$-$\amupitrack        & $\phantom{-}0.01$ & $0.05$ & $0.04$ \\ 
$-$\apid              & $-0.01$ & $0.02$ & $0.03$ \\      
$-$\amu               & $\phantom{-}0.03$ & $0.02$ & $0.02$ \\      
$-$\ahlt              & $\phantom{-}0.00$ & $0.01$ & $0.02$ \\ 
$-$\fbkg\Abkg         & $\phantom{-}0.02$ & $-$ & $0.03$  & $+$ \\  \hline  
$(1-\fbkg)\asls/2$ & $\phantom{-}0.16$ & 0.11 & 0.08 \\ 
$2/(1-\fbkg)$      & $\phantom{-}2.45$ &  $-$ & 0.18 & $\times$ \\ \hline
\asls           & $\phantom{-}$\aslsVal & \aslsStat & \aslsSyst  \\
  \end{tabular}
\end{table}

The previous analysis, based on $1.0\invfb$, used only candidates in the \phipi
region of the Dalitz plot, with different selection criteria, and used a
different fit method to determine the signal yields~\cite{LHCb-PAPER-2013-033}.
A more stringent selection resulted in a cleaner signal sample, but with
roughly 30\% fewer signal candidates in the \phipi region. As a cross check, the
approach of the previous analysis is repeated on the full $3.0\invfb$ data
sample and the result is compatible within one standard deviation.

\begin{figure}
  \centering
  \includegraphics[width=\figWidth\textwidth]{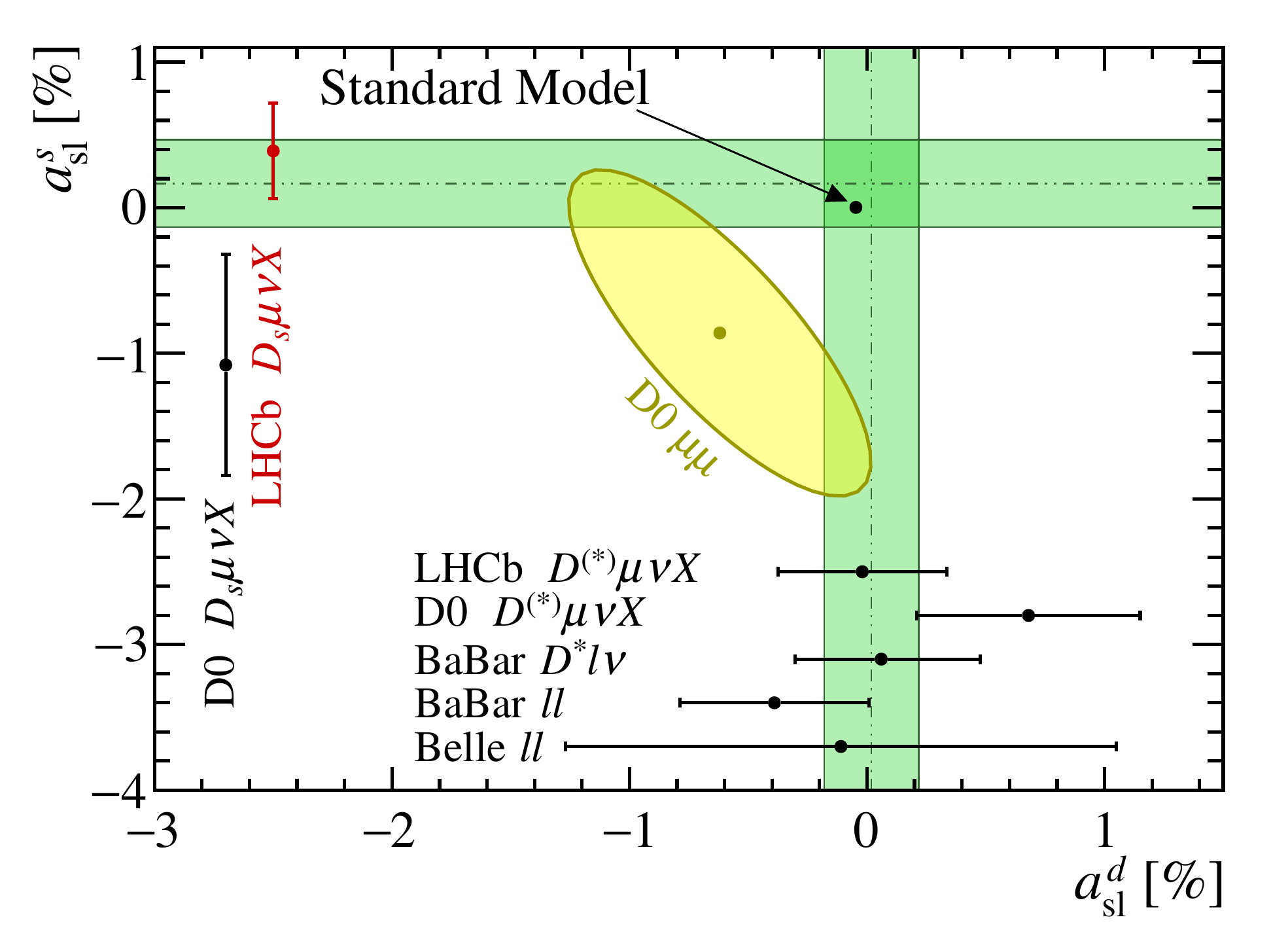}
  \caption{Overview of the most precise measurements of \asld and \asls. The
    horizontal and vertical bands indicate the naive averages of pure \asls and
    \asld measurements~\cite{Nakano:2005jb, Lees:2013sua, Lees:2014kep,
      Abazov:2012hha, LHCb-PAPER-2014-053, Abazov:2012zz}. The
    yellow ellipse represents the \dzero dimuon measurement with $\DGd/\Gd$ set
    to its SM expectation value~\cite{Abazov:2013uma}. The error bands and
    contours correspond to 68\% confidence level.}
  \label{fig:aslsasldplane}
\end{figure}

The twelve values of \asls for each Dalitz region, polarity and data-taking
period are consistent with each other. The combined result, taking into account 
all correlations, is
\begin{equation*}
  \asls = (\aslsVal \pm \aslsStat \pm \aslsSyst)\% \ , 
\end{equation*} 
where the first uncertainty is statistical, originating from the size of the
signal and calibration samples, and the second systematic. There is a small
correlation coefficient of $+0.13$ between this measurement and the \lhcb
measurement of \asld~\cite{LHCb-PAPER-2014-053}. The correlation mainly
originates from the muon detection asymmetry and from the effect of \asld, due
to \Bd background, on the measurement of \asls. Figure~\ref{fig:aslsasldplane}
displays an overview of the most precise measurements of \asld and
\asls~\cite{Nakano:2005jb, Lees:2013sua, Lees:2014kep, Abazov:2012hha,
  LHCb-PAPER-2014-053, Abazov:2012zz, Abazov:2013uma}. The simple averages of
pure \asl measurements, including the present \asls result and accounting for
the small correlation from \lhcb, are found to be $\asld = (0.02\pm0.20)\%$ and
$\asls=(0.17\pm0.30)\%$ with a correlation of $+0.07$. In combination, these two averages are marginally
compatible with the \dzero dimuon result ($p=0.5\%$) shown in Fig.~\ref{fig:aslsasldplane}.  
In summary, the determination of \asls presented in this letter is the
most precise to date. It shows no evidence for new physics effects and will serve to restrict models beyond the SM.

\section*{Acknowledgements}

We express our gratitude to our colleagues in the CERN
accelerator departments for the excellent performance of the LHC. We
thank the technical and administrative staff at the LHCb
institutes. We acknowledge support from CERN and from the national
agencies: CAPES, CNPq, FAPERJ and FINEP (Brazil); NSFC (China);
CNRS/IN2P3 (France); BMBF, DFG and MPG (Germany); INFN (Italy); 
FOM and NWO (The Netherlands); MNiSW and NCN (Poland); MEN/IFA (Romania); 
MinES and FANO (Russia); MinECo (Spain); SNSF and SER (Switzerland); 
NASU (Ukraine); STFC (United Kingdom); NSF (USA).
We acknowledge the computing resources that are provided by CERN, IN2P3 (France), KIT and DESY (Germany), INFN (Italy), SURF (The Netherlands), PIC (Spain), GridPP (United Kingdom), RRCKI and Yandex LLC (Russia), CSCS (Switzerland), IFIN-HH (Romania), CBPF (Brazil), PL-GRID (Poland) and OSC (USA). We are indebted to the communities behind the multiple open 
source software packages on which we depend.
Individual groups or members have received support from AvH Foundation (Germany),
EPLANET, Marie Sk\l{}odowska-Curie Actions and ERC (European Union), 
Conseil G\'{e}n\'{e}ral de Haute-Savoie, Labex ENIGMASS and OCEVU, 
R\'{e}gion Auvergne (France), RFBR and Yandex LLC (Russia), GVA, XuntaGal and GENCAT (Spain), Herchel Smith Fund, The Royal Society, Royal Commission for the Exhibition of 1851 and the Leverhulme Trust (United Kingdom).

\addcontentsline{toc}{section}{References}
\setboolean{inbibliography}{true}
\bibliographystyle{LHCb}
\bibliography{main,LHCb-PAPER,LHCb-CONF,LHCb-DP,LHCb-TDR,my-refs}

\newpage

\centerline{\large\bf LHCb collaboration}
\begin{flushleft}
\small
R.~Aaij$^{39}$,
B.~Adeva$^{38}$,
M.~Adinolfi$^{47}$,
Z.~Ajaltouni$^{5}$,
S.~Akar$^{6}$,
J.~Albrecht$^{10}$,
F.~Alessio$^{39}$,
M.~Alexander$^{52}$,
S.~Ali$^{42}$,
G.~Alkhazov$^{31}$
% ...

{\footnotesize \it
% ...
$ ^{5}$Clermont Universit{\'e}, Universit{\'e} Blaise Pascal, CNRS/IN2P3, LPC, Clermont-Ferrand, France\\
$ ^{6}$CPPM, Aix-Marseille Universit{\'e}, CNRS/IN2P3, Marseille, France\\
% ...
$ ^{10}$Fakult{\"a}t Physik, Technische Universit{\"a}t Dortmund, Dortmund, Germany\\
% ...
$ ^{31}$Petersburg Nuclear Physics Institute (PNPI), Gatchina, Russia\\
% ...
$ ^{38}$Universidad de Santiago de Compostela, Santiago de Compostela, Spain\\
$ ^{39}$European Organization for Nuclear Research (CERN), Geneva, Switzerland\\
% ...
$ ^{42}$Nikhef National Institute for Subatomic Physics, Amsterdam, The Netherlands\\
% ...
$ ^{47}$H.H. Wills Physics Laboratory, University of Bristol, Bristol, United Kingdom\\
% ...
$ ^{52}$School of Physics and Astronomy, University of Glasgow, Glasgow, United Kingdom\\
% ...
}
\end{flushleft}

\end{document}